\documentclass[10pt,journal,compsoc]{IEEEtran}

\ifCLASSOPTIONcompsoc
  \usepackage[nocompress]{cite}
\else
  \usepackage{cite}
\fi

\ifCLASSINFOpdf
\else
\fi
\usepackage{ulem}
\usepackage{times}
\usepackage{epsfig}
\usepackage{graphicx}
\usepackage{amsmath}
\usepackage{amssymb}
\usepackage{amsmath}
\DeclareMathOperator*{\argmin}{argmin}
\usepackage[pagebackref=true,breaklinks=true,letterpaper=true,colorlinks,bookmarks=false]{hyperref}

\begin{document}
%
\title{Unsupervised Learning of 3D Point Set Registration}

\author{Lingjing~Wang, ~Xiang~Li
        and~Yi~Fang
\thanks{L.Wang is with MMVC Lab, the Department of Mathematics, New York University, New York, NY, 30332 USA, e-mail: lingjing.wang@courant.nyu.edu. X.Li is with the MMVC Lab, New York University, New York, NY11201, USA, e-mail: xl1845@nyu.edu. Y.Fang is with MMVC Lab, Dept. of ECE, NYU Abu Dhabi, UAE and Dept. of ECE, NYU Tandon School of Engineering, USA, e-mail: yfang@nyu.edu.}
\thanks{Corresponding author. Email: yfang@nyu.edu}
}
\maketitle
\begin{abstract}
Point cloud registration is the process of aligning a pair of point sets via searching for a geometric transformation. Recent works leverage the power of deep learning for registering a pair of point sets. However, unfortunately, deep learning models often require a large number of ground truth labels for training. Moreover, for a pair of source and target point sets, existing deep learning mechanisms require explicitly designed encoders to extract both deep spatial features from unstructured point clouds and their spatial correlation representation, which is further fed to a decoder to regress the desired geometric transformation for point set alignment. To further enhance deep learning models for point set registration, this paper proposes Deep-3DAligner, a novel unsupervised registration framework based on a newly introduced deep \underline{S}patial \underline{C}orrelation \underline{R}epresentation (SCR) feature. The SCR feature describes the geometric essence of the spatial correlation between source and target point sets in an encoding-free manner. More specifically, our method starts with optimizing a randomly initialized latent SCR feature, which is then decoded to a geometric transformation (i.e., rotation and translation) to align source and target point sets. Our Deep-3DAligner jointly updates the SCR feature and weights of the transformation decoder towards the minimization of an unsupervised alignment loss. We conducted experiments on the ModelNet40 datasets to validate the performance of our unsupervised Deep-3DAligner for point set registration. The results demonstrated that, even without ground truth and any assumption of a direct correspondence between source and target point sets for training, our proposed approach achieved comparative performance compared to most recent supervised state-of-the-art approaches.
\end{abstract}


\section{Introduction}

\IEEEPARstart{P}{oint} set registration is a challenging but meaningful task, which has wide application in many fields \cite{myronenko2009image,ma2016non,wu2012online,klaus2006segment,maintz1998survey,raguram2008comparative,yuille1988computational,sonka2014image}. For example, point set registration algorithm can be used to align a pool of local frames to a global one for large-scale 3D reconstruction or 3D mapping \cite{Ding_2019_CVPR}.
\begin{figure*}
\begin{center}
\includegraphics[width=14cm]{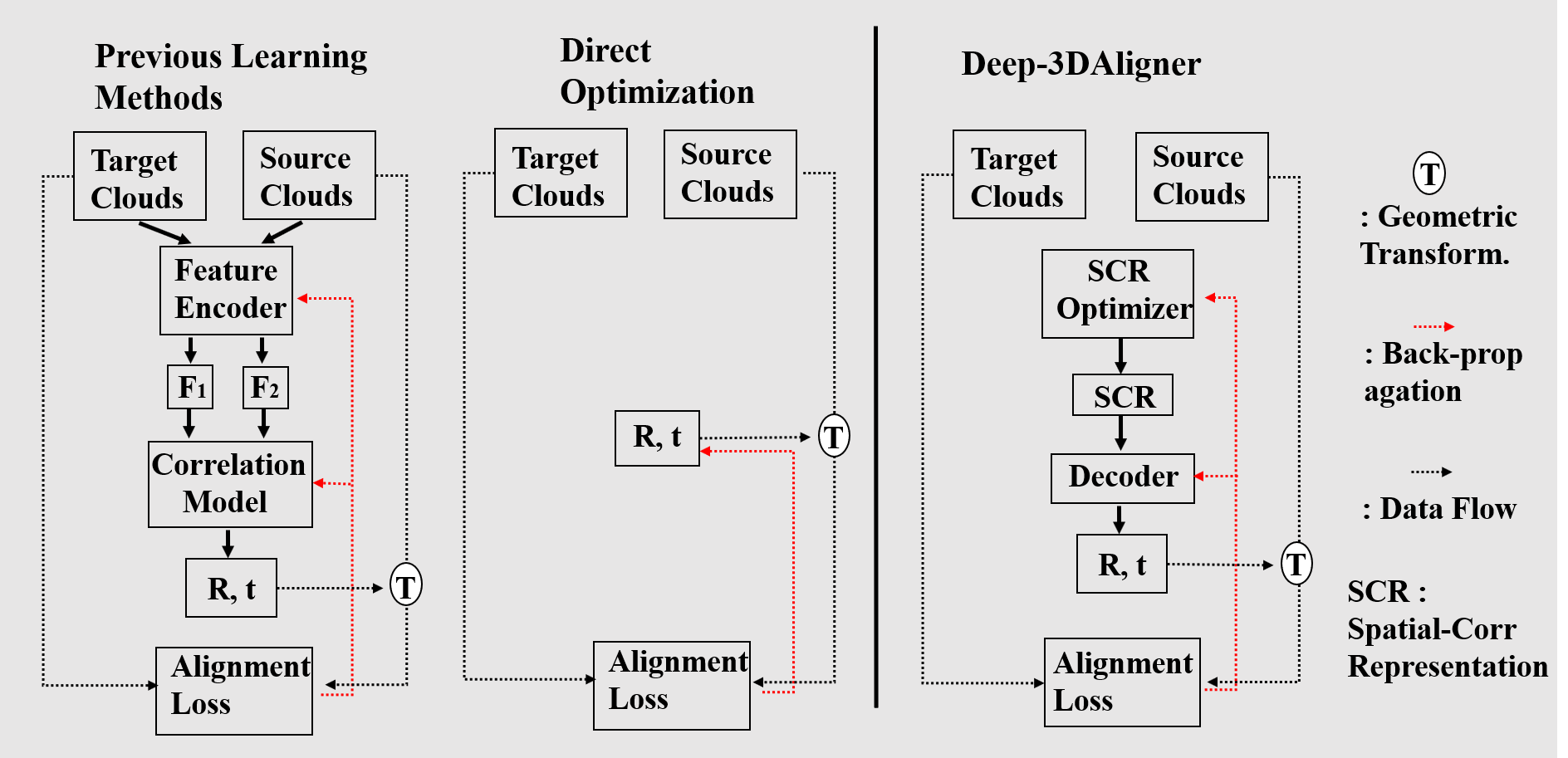}
\end{center}
\caption{Comparison of the pipeline between previous learning methods, direct optimization methods and our Deep-3DAligner for point set registration. Our method starts with optimizing a randomly initialized latent spatial correlation representation (SCR) feature, which is then decoded to the desired geometric transformation to align source and target point clouds, avoiding the explicit design of feature encoder and correlation module which is often challenging for point clouds input and increasing the model complexity by leveraging the structure of deep neural networks in comparison to direct optimization methods.}
\label{first}
\end{figure*}
Most existing non-learning methods solve the registration problem through an iterative optimization process to search the optimal geometric transformation to minimize a pre-defined alignment loss between transformed source point set and target point set \cite{myronenko2007non,ma2013robust,ma2014robust,ling2005deformation}. The geometric transformation can be modeled by a specific type of parametric transformation (e.g. rotation, translation, thin-plate spline and so on) \cite{besl1992method}. For example, one of the most commonly applied methods, iterative closest point (ICP) \cite{besl1992method}, estimates the rigid transformation based on a set of corresponding points. The ICP model, however, strongly depends on the initialization and has limited performance in choosing corresponding points. Moreover, iterative methods usually treat registration as an independent optimization process for each given pair of source and target point sets, which cannot transfer knowledge from registering one pair to another.

In recent years, deep-learning-based algorithms have been implemented in various industries and achieved great success, researchers are increasingly interested in bringing deep-learning-based solutions to the field of point set registration. As shown in Figure \ref{first}, instead of directly optimizing the transformation matrix towards a minimization of alignment loss in non-learning based methods, learning-based methods usually leverage modern feature extraction technologies for feature learning and then regress the transformation matrix based on the mutual information and correlation defined on the extracted features of source and target shapes. The most recent model, deep closest point (DCP) \cite{wang2019deep}, leverages DGCNN \cite{wang2019dynamic} for feature learning and a pointer network to perform soft matching. To refine the soft matching results to predict the final rigid transformation, the DCP model further proposes a singular value decomposition layer for fine-tuning. However, it is still challenging to design an explicit module for learning both the features from unstructured point clouds and their ``geometric relationship" \cite{Wang_2018_CVPR}. Existing works developed various models to compute the spatial correlation feature. For example, FlowNet3D \cite{liu2019flownet3d} tied to concatenate two global descriptors of source and target point sets; \cite{balakrishnan2018unsupervised} used a U-Net-based structure to mix the source and target volumetric shapes; \cite{rocco2017convolutional} proposed a correlation tensor calculated from source and target feature map and so on. In contrast, our paper proposes Deep-3DAligner, a novel unsupervised registration framework, as shown in Figure \ref{first}, relies on a directly optimizable SCR feature instead of requiring designing feature encoder and correlation module. Besides, Deep-3DAligner is trained in an unsupervised manner, which is different from the DCP that uses the ground-truth transformation parameters (i.e. rotation and translation matrix) for training.

With the development of the SCR feature, our proposed Deep-3DAligner framework is illustrated in Figure \ref{main}, which contains three main components. The first component is an SCR optimizer where the deep SCR feature is optimized from a randomly initialized feature. The second component is a transformation decoder which decodes the SCR feature to regress the transformation parameters for the point sets alignment. The third component is an alignment loss that measures the similarity between the transformed source point set and the target one. In the pipeline, there are two communication routes, indicated by black and red dashed lines. The communication route in black is for the data flow for the Deep-3DAligner paradigm, where the source and target point sets are used as input. The communication route in red is the back-propagation route with which the alignment loss is back-propagated to update the SCR and the transformation decoder. Our contribution is as follows:

\begin{itemize}

\item We introduce a novel unsupervised learning approach for the point set registration task.

\item We introduce a spatial correlation representation (SCR) feature which can eliminate the design challenges for encoding the spatial correlation between source and target point sets in comparison to learning-based methods. 

\item Experimental results demonstrate the effectiveness of the proposed method for point set registration, and even without ground truth transformation for training, our proposed approach achieved comparative performance compared to most recent supervised state-of-the-art approaches.

\end{itemize}




\section{Related Works}
\subsection{Iterative registration methods}
The development of optimization algorithms to estimate rigid and non-rigid geometric transformations in an iterative routine has attracted extensive research attention in past decades. Assuming that a pair of point sets are related by a rigid transformation, the standard approach is to estimate the best translation and rotation parameters in the iterative search routine, therein aiming to minimize a distance metric between two sets of points. The iterative closest point (ICP) algorithm \cite{besl1992method} is one successful solution for rigid registration. It initializes an estimation of a rigid function and then iteratively chooses corresponding points to refine the transformation. However, the ICP algorithm is reported to be vulnerable to the selection of corresponding points for initial transformation estimation. Go-ICP \cite{yang2015go} was further proposed by Yang et al. to leverage the BnB scheme for searching the entire 3D motion space to solve the local initialization problem brought by ICP. Zhou et al. proposed fast global registration \cite{zhou2016fast} for the registration of partially overlapping 3D surfaces. The TPS-RSM algorithm was proposed by Chui and Rangarajan \cite{chui2000new} to estimate parameters of non-rigid transformations with a penalty on second-order derivatives. As a classical non-parametric method, coherence point drift (CPD) was proposed by Myronenko et al. \cite{myronenko2007non}, which successfully introduced a process of fitting the Gaussian mixture likelihood to align the source point set with the target point set. Existing classical algorithms have achieved great success on the registration task. Although the independent iterative optimization process limits the efficiency of registering a large number of pairs, inspiring us to design a learning-based system for this task.

\subsection{Learning-based registration methods} In recent years, learning-based methods have achieved great success in many fields of computer vision \cite{su2015multi,sharma2016vconv,maturana2015voxnet,qi2017pointnet,verma2018feastnet,masci2015geodesic,zeng20173dmatch,wang20173densinet,wang2018unsupervised,wang2020few}. In particular, recent works have started a trend of directly learning geometric features from cloud points (especially 3D points), which motivates us to approach the point set registration problem using deep neural networks \cite{rocco2017convolutional,balakrishnan2018unsupervised,zeng20173dmatch,qi2017pointnet,verma2018feastnet,masci2015geodesic,chen2019arbicon,wang2019non,wang2019coherent,li2019pc}. PointNetLK \cite{aoki2019pointnetlk} was proposed by Aoki et al. to leverage the newly proposed PointNet algorithm for directly extracting features from the point cloud with the classical Lucas $\&$ Kanade algorithm for the rigid registration of 3D point sets. Liu et al. proposed FlowNet3D \cite{liu2019flownet3d} to treat 3D point cloud registration as a motion process between points. Wang et al. proposed a deep closest point \cite{wang2019deep} model, which first leverages the DGCNN structure to exact the features from point sets and then regress the desired transformation based on it. Balakrishnan et al. \cite{balakrishnan2018unsupervised} proposed a voxelMorph CNN architecture to learn the registration field to align two volumetric medical images. For the registration of 2D images, an outstanding registration model was proposed by Rocco et al. \cite{rocco2017convolutional}. For the learning-based registration solutions listed above, the main challenge concerns how to effectively model the ``geometric relationship" between source and target objects in a learning-based approach. For example, \cite{rocco2017convolutional} proposed a correlation tensor between the feature maps of source and target images. \cite{balakrishnan2018unsupervised} leveraged a U-Net-based structure to concatenate features of source and target voxels. \cite{liu2019flownet3d}
\cite{aoki2019pointnetlk} used a PointNet-based structure, and \cite{wang2019deep} used a DGCNN structure to learn the features from a point set for further registration decoding.
In contrast, we first propose a model-free structure to skip the encoding step. Instead, we initialize an SCR feature without pre-defining a model, which is to be optimized with the weights of the network from the alignment loss back-propagation process.
\begin{figure*}
\begin{center}
\includegraphics[width=17cm]{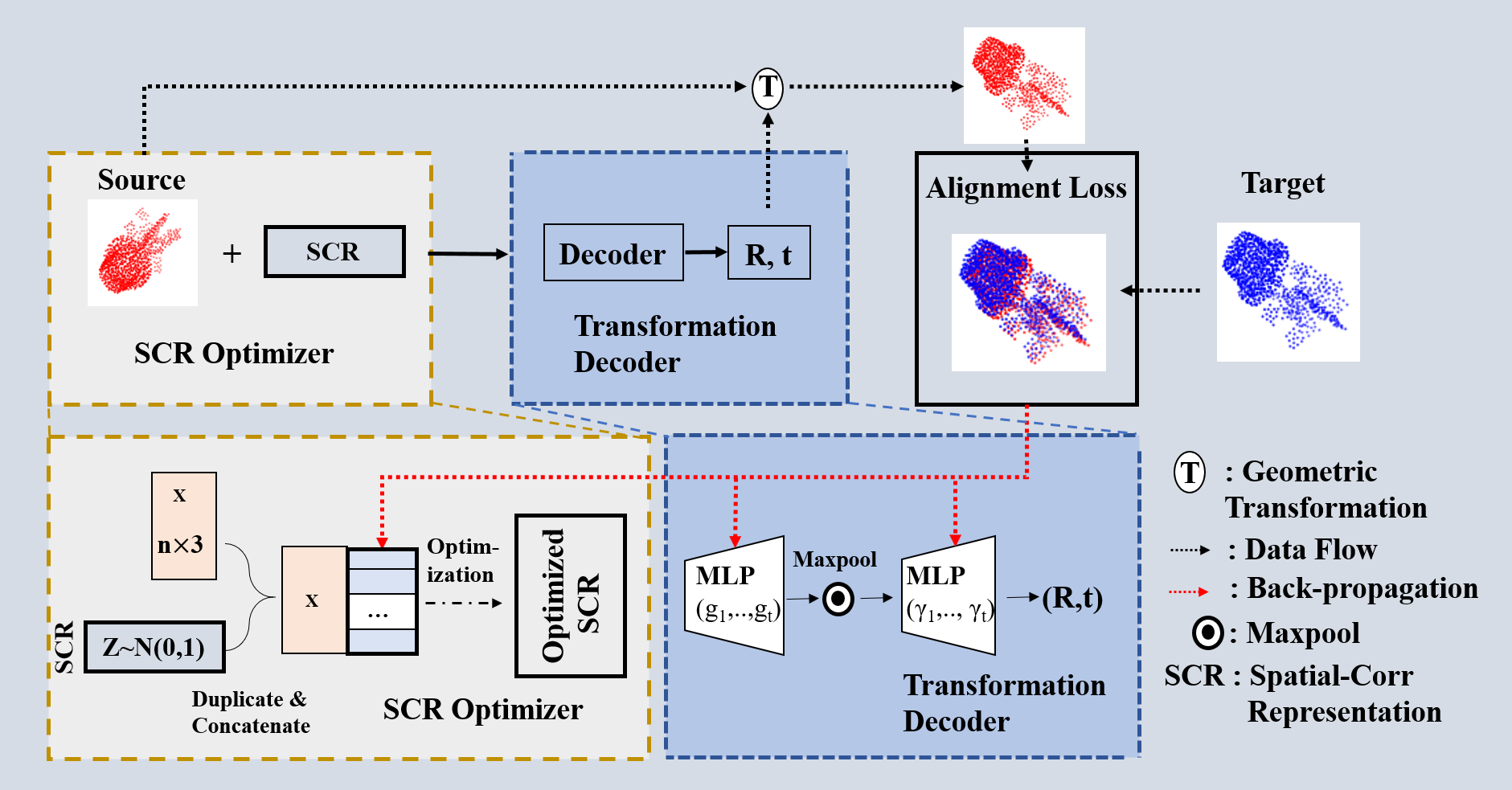}
\end{center}
\caption{Our pipeline. For a pair of input source and target point sets, our method starts with the SCR optimization process to generate a spatial-correlation representation feature, and a transformation regression process further decodes the SCR feature to the desired geometric transformation. The alignment loss is back-propagated to update the weight of the transformation decoder and the SCR feature during the training process. For testing, the weights of the transformation decoder remain constantly without updating.}
\label{main}
\end{figure*}

\section{Approach}
We introduce our approach in the following sections. First, we define the learning-based registration problem in section \ref{sc_problem}. In section \ref{sc_scr}, we introduce our spatial-correlation representation. The transformation decoder is illustrated in section \ref{sc_trans_reg}. In section \ref{sc_loss}, we provide the definition of the loss function. Section \ref{sc_optim} illustrates the newly defined optimization strategy. 

\subsection{Problem statement}\label{sc_problem}
Given training dataset $\bold{D}=\{(\bold{S_i}, \bold{G_j}) \text{ ,where } \bold{S_i}, \bold{G_j} \subset \mathbb{R}^N (N=2 \text{ or } N=3) \}$, the optimization task of a deep-learning-based method for registration problem can be generally defined in the following way. We assume the existence of a function $g_{\theta}(\bold{S_i},\bold{G_j}) = \phi$ using a neural network structure, where $\phi$ represents the parameters of the network. The rigid point set registration is represented by a homogeneous transformation matrix, which is composed by a rotation matrix $\bold{R} \in SO(3)$ and a translation vector $\bold{t} \in \mathbb{R}^3$. Given a pair of input source and target point sets $(\bold{S_i},\bold{G_j})$, a trained model is able to predict the parameters $\phi$ based on the optimized weights $\bold{\theta^{optimal}}$ in the neural network structure. A pre-defined alignment metric between transformed source and target point sets can be defined as objective loss function to update weights $\theta$. For a given dataset $\bold{D}$, a stochastic gradient-descent-based algorithm can usually be utilized for the optimization of the weights $\theta$ to minimize the pre-defined loss function:
\begin{equation}
\begin{split}
\bold{\theta^{optimal}} =\argmin_{\theta}[\mathbb{E}_{(\bold{S_i},\bold{G_j})\sim \bold{D}}[\mathcal{L}(\bold{S_i},\bold{G_j}, g_{\theta}(\bold{S_i},\bold{G_j}))]],
\end{split}
\end{equation}
where $\mathcal{L}$ represents a similarity metric. 

\subsection{Spatial-Correlation Representation}\label{sc_scr}
In this paper, we define the spatial correlation representation as the latent feature that characterizes the essence of spatial correlation between a given pair of source and target point sets. As shown in Figure \ref{first}, to compute the SCR feature, source and target point sets are usually fed to a feature in previous works (i.e. PointNet \cite{qi2017pointnet}) for the deep spatial feature extraction, and followed with a pre-defined correlation module (i.e. \cite{rocco2017convolutional}). However, the design of an appropriate feature encoder for unstructured point clouds is challenging compared to the standard discrete convolutions assume the availability of a grid structured input (e.g. 2D image). Furthermore, the design of a correlation module for a pair of input spatial features has a significant impact on the transformation decoder. The limitation of the hand-crafted design of modules for the extraction of individual spatial feature and spatial correlation feature motivates us to design a model-free based SCR as described below. 

To eliminate the side effects of the hand-craft design in feature encoder and correlation module, as shown in Figure \ref{main}, we define a trainable latent SCR (Spatial-Correlation Representation) feature for each pair of point sets. As shown in Figure \ref{first}, our SCR optimizer, which takes input as a randomly initialized latent vector to reason the optimal SCR for the point set registration task, replaces the previously hand-crafted feature encoder and correlation module. More specifically, as shown in Figure \ref{main}, for a pair of source and target point sets $(\mathbf{S}, \mathbf{G})$, the randomly initialized latent vector $\bold{z}$ from Gaussian distribution as an initialized SCR. The initialized SCR is optimized during the training process together with the transformation decoder. In this way, in comparison with previous methods, we avoid the challenging problem of explicit defining the spatial correlation between two point sets but simply optimize the SCR feature from a random initialized status during the model training process. The implicit design of SCR allows Deep-3DAligner more flexibility in spatial correlation feature learning that is more adaptive for the alignment of unseen point sets.

\subsection{Transformation Decoder}\label{sc_trans_reg}
Given the above spatial-correlation representation (SCR) feature, we then design a decoding network to regress the desired transformation parameters, as illustrated in Figure \ref{main}. More specifically, we first formulate the input by stacking the coordinates of each point $x$ in source point set $S$ with its corresponding SCR feature $\bold{z}$. We note this input as $[x, \mathbf{z}]$. Then we define a multi-layer perceptron (MLP) architecture for learning the parameters of the rigid transformation to transform the source point set toward the target point set. This architecture includes successive MLP layers with the ReLU activation function, $\{g_i\}_{i=1,2,...,s}$, such that $g_i : \mathbb{R}^{v_{i}}\to \mathbb{R}^{v_{i+1}}$, where $v_{i}$ and $v_{i+1}$ are the dimensions of the layer inputs and outputs respectively. Then, we use a max pool layer to exact the global feature $\mathbf{L}$, calculated as:
\begin{equation}
\begin{split}
\bold{L}=Maxpool\{g_sg_{s-1}...g_1([\bold{x_i},\bold{z}])\}_{\bold{x_i}\in \bold{S}}
\end{split}
\end{equation}
where the notation [*,*] represents the concatenation of vectors in the same domain. 

We further decode the global feature $\bold{L}$ to the transformation parameters by a second network, which includes $t$ successive MLP layers with a ReLU activation function $\{\gamma_i\}_{i=1,2,...,t}$ such that $\gamma_i : \mathbb{R}^{w_{i}}\to \mathbb{R}^{w_{i+1}}$, where $w_{i}$ and $w_{i+1}$ are the dimensions of the layer inputs and outputs respectively. 
\begin{equation}
\begin{split}
\bold{\mathbf{\phi}}=\gamma_t\gamma_{t-1}...\gamma_1(\mathbf{L})
\end{split}
\end{equation}
Now, we can get the transformed source point set $\bold{S'}$ by
\begin{equation}
\begin{split}
\bold{S'} =\mathbf{\bold{T}_{\phi}}(\bold{S})
\end{split}
\end{equation}
where $\bold{T}_{\phi}$ denotes the transformation function defined by the predicted transformation parameters $\phi$. Based on the transformed source point set and the target point set, we can further define the alignment loss function in the next section.

\subsection{Loss function}\label{sc_loss}
In our unsupervised setting, we do not have the ground truth transformation for supervision and we do not assume a direct correspondence between these two point sets. Therefore, a distance metric between two point sets, instead of the point/pixel-wise loss is desired. In addition, A suitable metric should be differentiable and efficient to compute. In this paper, we adopt the Chamfer distance proposed in \cite{fan2017point} as our loss function. The Chamfer loss is a simple and effective alignment metric defined on two non-corresponding point sets. We formulate the Chamfer loss between our transformed source point set $T_{\phi}(\mathbf{S})$ and target points set $\mathbf{G} $ as:
\begin{equation} 
\begin{split}L_{\text{Chamfer}}(T_{\phi}(\mathbf{S}),\mathbf{G})
 &= \sum_{x\in T_{\phi}(\mathbf{S})}\min_{y \in \mathbf{G}}||x-y||^2_2\\
 &+ \sum_{y\in \mathbf{G}}\min_{x \in T_{\phi}(\mathbf{S})}||x-y||^2_2
\end{split}
\end{equation}
where $\phi$ represents all the parameters to define the transformation and $\phi$ is predicted from our model based on the optimized SRC feature and the trained decoder.

\subsection{Optimization Strategy}\label{sc_optim}

In section \ref{sc_scr}, we define a set of trainable latent vectors $\bold{z}$, one for each pair of point sets as the SCR feature. During the training process, these latent vectors are optimized along with the weights of network decoder using a stochastic gradient descent-based algorithm. For a given training dataset $\bold{D}$, our training process can be expressed as:
\begin{equation}
\begin{split}
\bold{\theta^{optimal}, z^{optimal}} =\argmin_{\theta, \mathbf{z}}[\mathbb{E}_{(\bold{S_i},\bold{G_j})\sim \bold{D}}[\mathcal{L}(\bold{S_i},\bold{G_j}, g_{\theta}(\bold{S_i},\mathbf{z}))]],
\end{split}
\end{equation}
where $\mathcal{L}$ represents the pre-defined loss function.

For a given testing dataset $\bold{W}$, we fix the network parameters $\tilde{\theta}= \bold{\theta^{optimal}}$ and only optimize the SRC features:
\begin{equation}
\begin{split}
\bold{z^{optimal}} =\argmin_{\mathbf{z}}[\mathbb{E}_{(\bold{S_i},\bold{G_j})\sim \bold{W}}[\mathcal{L}(\bold{S_i},\bold{G_j}, g_{\tilde{\theta}}(\bold{S_i},\mathbf{z}))]].
\end{split}
\end{equation}
The learned decoder network parameters $\tilde{\theta}$ here provides a prior knowledge for the optimization of SRC. After this optimization process, the desired transformation can be determined by $T_{\phi}=T_{g_{\tilde{\theta}}(\bold{S_i},\mathbf{z_i^{optimal}})}$ and the transformed source shape can be generated by $\mathbf{S_i'}=T_{\phi}(\mathbf{S_i}),  \forall \bold{S_i} \in \bold{W}$. 


\begin{figure*}
\begin{center}
\includegraphics[width=17cm]{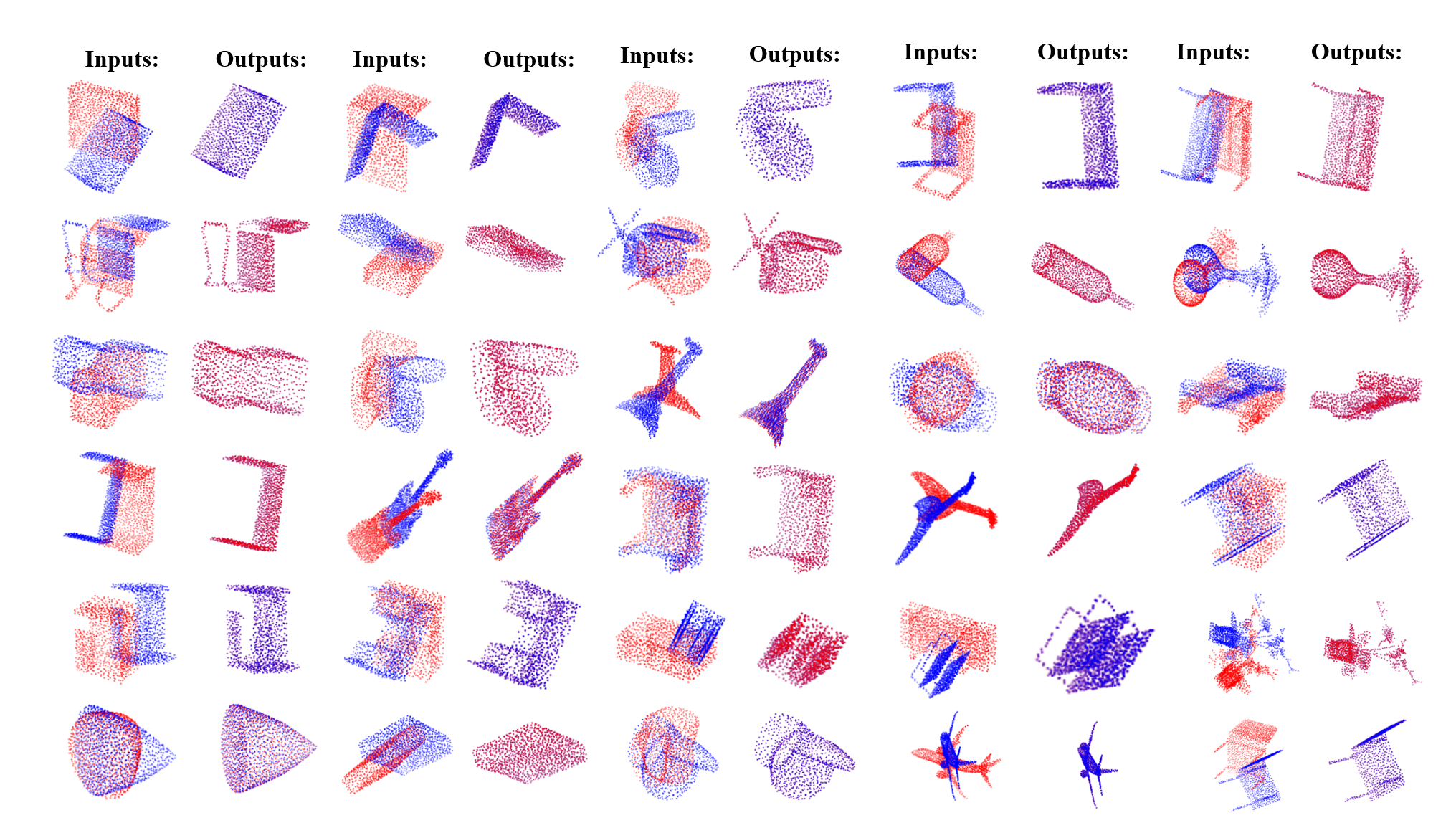}
\end{center}
\caption{Randomly selected qualitative results of our model for registration of unseen samples. Left columns: inputs. Right columns: outputs. The red points represent source point sets, and the blue points represent the target point sets.}
\label{allres}
\end{figure*}

\begin{table*}
\small
\begin{center}
 \resizebox{\textwidth}{!}{\begin{tabular}{ccccccc}
\hline
Model& MSE(R) &RMSE(R)& MAE(R) &MSE(t)& RMSE(t)& MAE(t)
\\
 \hline
ICP \cite{besl1992method}& 894.897339& 29.914835& 23.544817& 0.084643 &0.290935& 0.248755\\
Go-ICP \cite{yang2015go}& 140.477325& 11.852313& 2.588463& 0.000659& 0.025665& 0.007092\\
FGR \cite{zhou2016fast}& 87.661491& 9.362772& 1.999290& 0.000194& 0.013939 &0.002839\\
PointNetLK \cite{aoki2019pointnetlk}& 227.870331& 15.095374& 4.225304& 0.000487& 0.022065& 0.005404\\
DCPv1+MLP(Supervised)\cite{wang2019deep} & 21.115917& 4.595206&  3.291298& 0.000861 &0.029343&0.022501\\
\hline
DCPv2+MLP(Supervised)\cite{wang2019deep} & 9.923701&  3.150191& 2.007210&0.000025&  0.005039& 0.003703\\
DCPv1+SVD(Supervised)\cite{wang2019deep} & 6.480572& 2.545697& 1.505548& 0.000003 &0.001763& 0.001451\\
DCPv2+SVD(Supervised)\cite{wang2019deep} & 1.307329& 1.143385& \textbf{0.770573}& \textbf{0.000003}& \textbf{0.001786}& \textbf{0.001195}\\
\hline
Deep-3DAligner (MLP-based, Unsupervised) & \textbf{1.154405} &\textbf{1.074432}&0.830864&0.000444 &0.020904& 0.014533\\
\hline
\end{tabular}}
\end{center}
\caption{ModelNet40: Test on unseen point clouds. Our model is trained in an unsupervised manner without  any ground-truth labels. Our model does not require attention mechanism and SVD-based fine-tuning processes.}
\label{ttt2}
\end{table*}
\section{Experiments}\label{sc_exp}
We describe the experimental dataset and settings in section \ref{sc_dataset} and section \ref{sc_setting} respectively. In sections \ref{sc_exp1} and \ref{sc_exp2}, we test our model's performance on different settings and compare our model's performance with state-of-the-art methods. In section \ref{sc_pd}, we further demonstrate the robustness of our model in the presence of P.D. noise. In section \ref{sc_di}, we further demonstrate the robustness of our model in the presence of D.I. noise. In section \ref{sc_do}, we further demonstrate the robustness of our model in the presence of D.O. noise. In section \ref{sc_direct}, we compare our model's performance with the direct optimization version.

\subsection{Dataset preparation}\label{sc_dataset}
We test the performance of our model for 3D point set registration on the ModelNet40 dataset. This dataset contains 12311 pre-processed CAD models from 40 categories. For each 3D point object, we uniformly sample 1024 points from its surface. Following the settings of previous work, points are centered and re-scaled to fit in the unit sphere. To demonstrate the robustness of our model in the presence of various noise types, we add noise to the target point sets for model evaluation. To prepare the position drift (P.D.) noise, a zero-mean Gaussian is applied to each point in the target point set. The level of P.D. noise is defined as the standard deviation of Gaussian Noise. To prepare the data incompleteness (D.I.) noise, we randomly remove a certain amount of points from the entire point set. The level of D.I. noise is defined as the ratio of the eliminated points and the entire set. To prepare the data outlier (D.O.) noise, we randomly add a certain amount of points generated by a zero-mean Gaussian to the point set. The level of D.O. noise is defined as the ratio of the added points to the entire point set. 
\begin{table*}
\small
\begin{center}
 \resizebox{\textwidth}{!}{\begin{tabular}{ccccccc}
\hline
Model& MSE(R) &RMSE(R)& MAE(R) &MSE(t)& RMSE(t)& MAE(t)
\\
 \hline
ICP\cite{besl1992method}& 892.601135&29.876431&23.626110&0.086005&0.293266&0.251916\\
Go-ICP \cite{yang2015go}& 192.258636&13.865736&2.914169&0.000491&0.022154&0.006219\\
FGR \cite{zhou2016fast}&97.002747&9.848997&1.445460&0.000182&0.013503&0.002231\\
PointNetLK \cite{aoki2019pointnetlk}& 306.323975&17.502113&5.280545&0.000784&0.028007&0.007203\\
DCPv1+SVD (Supervised) \cite{wang2019deep} &  19.201385&4.381938&2.680408&0.000025&0.004950&0.003597\\
DCPv2+SVD (Supervised) \cite{wang2019deep} & 9.923701& 3.150191& 2.007210& \textbf{0.000025} &\textbf{0.005039}&\textbf{0.003703}\\
Deep-3DAligner (MLP-based, Unsupervised) & \textbf{3.715267} &\textbf{1.485832}&\textbf{1.040233}&0.000822 &0.026767& 0.022763\\
\hline
\end{tabular}}
\end{center}
\caption{ModelNet40: Test on unseen categories. Our model is trained in an unsupervised manner without ground-truth labels. Our model does not require SVD-based fine-tuning processes.}
\label{ttt5}
\end{table*}
\subsection{Experimental Settings}\label{sc_setting}
We train our network using batch data from the training data set $\{(\mathbf{S_i},\mathbf{G_i}) | \mathbf{S_i}, \mathbf{G_i} \in \mathbf{D} \}_{i=1,2,...,b}$. We set the batch size $b$ to 128. The latent vectors are initialized from a Gaussian distribution $\mathcal{N}(0,1)$ with a dimension of 2048. For the decoding network, the first part includes 2 MLP layers with dimensions (256,128) and a max pool layer. Then, we use 3 additional MLPs with dimensions of (128, 64, 3) for decoding the rotation matrix and with dimensions of (128, 64, 3) for decoding the translation matrix. We use the leaky-ReLU \cite{xu2015empirical} activation function and implement batch normalization \cite{ioffe2015batch} for every layer except the output layer. Our model is optimized with Adam optimizer. The learning rate is set as 0.001 with exponential decay of 0.995 at each epoch. For the outlier and missing point cases, we clip the Chamfer distance by a fixed value of 0.1.

We use the mean squared error (MSE), root mean squared error (RMSE), and mean absolute error (MAE) to measure the performance of our model and all comparing methods. Lower values indicate better alignment performance. All angular measurements in our results are in units of degrees. The ground-truth labels are only used for the performance evaluation and are not used during the training/testing process. 

For performance evaluation, we compare our method with both supervised and unsupervised methods. The current state-of-the-art results are achieved by DCP, which is a supervised approach including four different versions. The version DCPv1+MLP uses deep neural networks (DNNs) to model the transformation. The version DCPv2+MLP improves its performance by integrating a supervised attention mechanism. DCPv1+SVD further improves its performance by integrating an additional SVD-based fine-tuning process, and DCPv2+SVD integrates both attention and SVD to further boost the performance. Since our Deep-3DAligner is an unsupervised approach and only use DNNs as auxiliary function to model the transformation, we will use DCPv1+MLP as the baseline model for compassion. 

\subsection{Full Dataset Training \& Testing}\label{sc_exp1}
In this experiment, we follow previous works, DCP \cite{wang2019deep} and PointNetLK \cite{aoki2019pointnetlk}, to test our model for 3D point set registration on unseen point sets.\\

\noindent{\textbf{Experiment Setting:}}
For the 12,311 CAD models from the ModelNet40, following exactly DCP's setting, we split the dataset into 9,843 models for training and 2,468 models for testing. As in DCP, we treat each 3D shape as our source point set, and randomly apply a rigid transformation on it along each axis to generate our target point set. The rotation is uniformly sampled from 0 to 45 degrees, and the translations are uniformly sampled in $[-0.5, 0.5]$. Note that we follow exactly the same experimental setting as the previous work in DCP for synthetic data simulation, where both source and target point sets are simulated with the same sampling. We train our Deep-3DAligner, DCP and PointNetLK on the divided training dataset and then evaluate the performance on the testing set. ICP, Go-ICP, and FGR are tested directly on the testing dataset. Note that our model is trained without using any ground-truth information, and our model does not require the SVD-based fine-tuning processes as used in DCP.\\

\noindent{\textbf{Results:}}
We list the quantitative experimental results in Table \ref{ttt2}. In this table, we evaluate the performance based on the prediction errors of rotation angles and translation vectors. The first three columns illustrate the comparison results for the rotation angle prediction. As we can see from the results, our method achieves significantly better performance than the baseline DCPv1+MLP model and also get even slightly better or comparative performance against the state-of-the-art approach (DCPv2+SVD). For example, our method achieves 1.074 RMSE(R) in rotation prediction, compared to 1.143(R) achieved by DCPv2+SVD. The last three columns illustrate the comparison results for translation prediction. As we can see from the results, our method achieves slightly better performance against DCPv1+MLP. One may note that the translation vector prediction performance of our model is inferior to that of DCPv2+MLP, DCPv1+SVD, DCPv2+SVD. The reason for this gap is that DCPv2 +MPL/SVD adopts an additional attention mechanism in its network for enhancement. DCPv1/DCPv2+SVD leverage SVD as an additional fine-tuning process to refine their results. SVD results in additional computational complexity, whereas our method uses only MLP-based networks and is trained end to end. Moreover, DCP assumes the same sampling of points and we tested that DCP experienced a severe performance degradation for randomly sampled points of source and target shapes, whereas our model with Chamfer distance is robust to the way of point sampling. As an unsupervised learning paradigm, we do not use ground-truth labels or any correspondence relationship between source and target points for training.  Chamfer distance-based loss is less sensitive to the translation, which possibly contributes to the deficit of translation prediction performance. We will explore other loss functions in a future study to address this problem. We note that the result of translation prediction is still better than PointNetLK. As shown in Figure \ref{allres}, we randomly select the qualitative results from the testing dataset. For shapes in various poses, the registration results indicate our model achieves remarkable performance in aligning the source point sets with the target point sets.

\subsection{Category Split}\label{sc_exp2}
In this experiment, we follow previous works, DCP and PointNetLK, to test our model for 3D point set registration on objects from unseen categories. \\
\begin{table*}
\centering
\small
\begin{tabular}{ccccccc}
\hline
Model & MSE(R)& MAE(R) & MSE(t) & MAE(t)\\
\hline
Ours & 2.9240& 0.8541 & 0.0002 & 0.012  \\
DCP&12.040397&2.190995&0.000642&0.015552\\
\hline
\end{tabular}
\caption{Quantitative result for 3D point set registration in presence of P.D. noise.}
\label{tab777}
\end{table*}

\begin{figure*}
\begin{center}
\includegraphics[width=17cm]{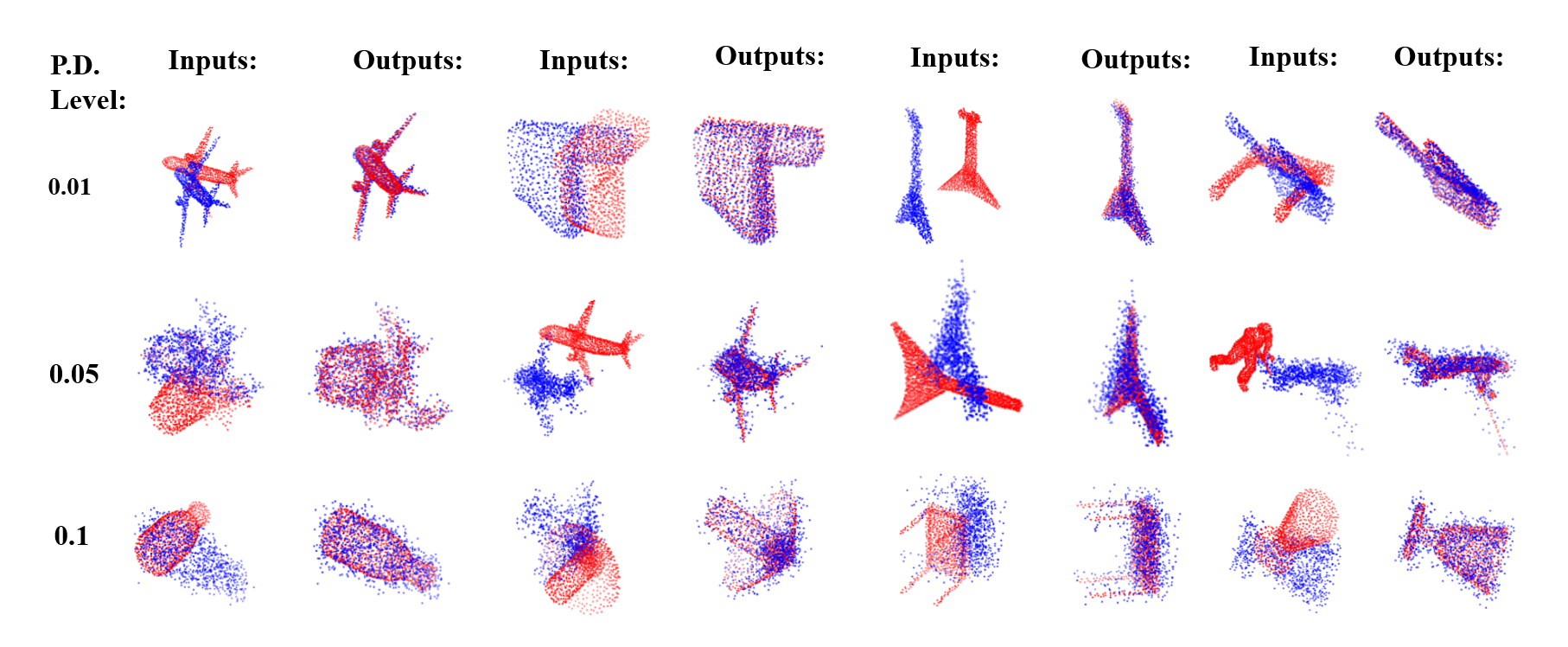}
\end{center}
\caption{Randomly selected qualitative results in presence of P.D. noise. Left columns: inputs. Right columns: outputs. The red points represent source point sets, and the blue points represent the target point sets.}
\label{fig5}
\end{figure*}
\noindent{\textbf{Experiment Setting:}}
To test the generalizability of our model, we split ModelNet40 evenly by category into training and testing sets in the same way as DCP. Our Deep-3DAligner, DCP, and PointNetLK are trained on the first 20 categories and test on the remaining categories. ICP, Go-ICP, and FGR are also tested on the held-out categories. Note that we follow the exact same experimental setting as the previous work in DCP for synthetic data simulation, where both source and target point sets are simulated with the same sampling. Our model is trained without using any ground-truth information, and our model does not require any SVD-based fine-tuning processes.\\

\noindent{\textbf{Results:}} As shown in Table \ref{ttt5}, the quantitative results indicate that our model achieves superior generalization ability on unseen categories as an unsupervised method. In comparison, all the supervised learning methods experienced a dramatic performance drop compared to the results in Table \ref{ttt2}. For example, we can see that PointNetLK and DCPv2+SVD obtain a MSE(R) of 227.87 and 1.31 in ``the training/testing split test''  as described in section \ref{sc_exp1} (see Table \ref{ttt2}). However, the corresponding values in ``seen/unseen categories test'' as described in this section increase to 306.32 and 9.92 respectively (see Table \ref{ttt5}). The MSE(R) of PointNetLK increased from 227.87 for unseen point clouds to 306.324 for unseen categories. Unsupervised algorithms, such as ICP and FGR, achieve similar accuracy for unseen categories and unseen point clouds. Our method has a small performance drop for the unseen categories compared to the results for unseen point clouds. Particularly, in the prediction of the rotation matrix for unseen categories, our method outperforms state-of-the-art DCPv2-SVD by a large margin (6.21) in MSE(R). 

\subsection{Resistance to Point Drifts (P.D.) Noise}\label{sc_pd}
In this experiment, we further verify our model's performance for 3D rigid point set registration in the presence of P.D. noise.\\

\noindent{\textbf{Experiment Setting:}} To test our model's performance for 3D rigid point set registration in the presence of P.D. noise, we firstly split ModelNet40 as explained in section 4.3. Then we add the P.D. noise on the target shape by the way introduced in section 4.1. In this section, we choose the DCP as the baseline model for comparison. We train the DCP and our model using our prepared training dataset and then test them using the testing dataset. The quantitative results with a P.D. noise level of 0.01 are demonstrated in Table \ref{tab777} and additional random selected qualitative results for various P.D. noise levels from 0.01 to 0.1 are shown in Figure \ref{fig5}. Note that we follow exactly the source code provided by DCP and the default settings in DCP for training and testing their model. Our model is trained without using any ground-truth information, and our model does not require any SVD-based fine-tuning processes.\\
\begin{table*}
\centering
\small
\begin{tabular}{ccccccc}
\hline
Model & MSE(R)& MAE(R) & MSE(t) & MAE(t)\\
\hline
Ours & 7.3354&1.4702 & 0.0008 &0.0222  \\
DCP & 34.624447&3.720148 & 0.002301 &0.032245  \\
\hline
\end{tabular}
\caption{Quantitative result for 3D point set registration in presence of D.I. noise.}
\label{tab888}
\end{table*}
\begin{figure*}
\begin{center}
\includegraphics[width=17cm]{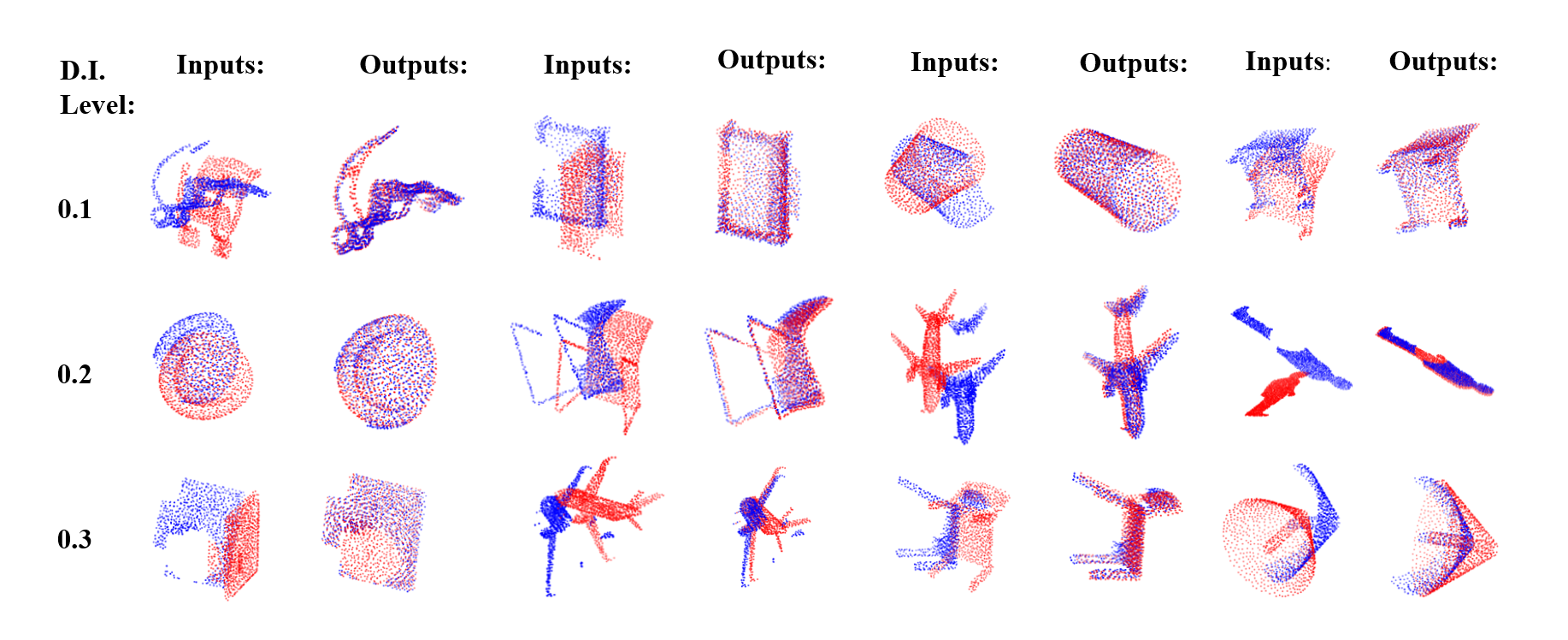}
\end{center}
\caption{Randomly selected qualitative results on dataset in presence of D.I. noise. Left columns: inputs. Right columns: outputs. The red points represent source point sets, and the blue points represent the target point sets.}
\label{fig6}
\end{figure*}

\noindent{\textbf{Results:}} As shown in Table \ref{tab777}, our method is more robust to P.D. noise. In comparison to DCP, for all the shown metrics of alignment performance our method achieves much better results than DCP. Especially for the rotation matrix estimation, our method achieves an MSE of 2.92 in comparison to 12.04 achieved by DCP. In addition, from the qualitative results shown in Figure \ref{fig5}, we notice that when P.D. noise level even increases from 0.01 to 0.1, the alignment result is nearly perfect for most cases.

\subsection{Resistance to Data Incompleteness (D.I.) Noise}\label{sc_di}
In this experiment, we further verify our model's performance for 3D rigid point set registration in the presence of D.I. noise.\\

\noindent{\textbf{Experiment Setting:}} To test our model's performance for 3D rigid point set registration in the presence of D.I. noise, we firstly split ModelNet40 as explained in section 4.3. Then we add the D.I. noise on the target shape by the way introduced in section 4.1. We compare our model with the state-of-the-art DCP model. Both models are trained using the training set and then tested on the test set. The quantitative result for D.I. noise level 0.1 is demonstrated in Table \ref{tab888}. Figure \ref{fig6} gives some randomly selected registration results for various D.I. noise levels from 0.1 to 0.3. One should note that our model is trained without using any ground-truth information, and our model does not require any SVD-based fine-tuning processes. We adjust the Chamfer distance loss by clipping large values which exceed 0.1.\\

\noindent{\textbf{Results:}} As shown in Table \ref{tab888}, for dataset in presence of D.I. noise, our model shows superior performance in comparison to DCP. Moreover, our proposed model achieves better performance than DCP model on all evaluation metrics. For the precision of rotation matrix estimation, our method achieves an MSE(R) of 7.33 in comparison to 36.62 achieved by DCP. However, as can be seen in Figure \ref{fig6}, our model performs well when the D.I. level is lower than 0.2 but experiences a performance drop when D.I. level increased to 0.3. When the D.I noise level increases to 0.3, our model performs well on objects with the missing parts in the middle of the body. For example, the cases such as the screen case in the first column and the chair case in the third column, driven by the Chamfer distances loss, the optimal alignment solution is correct.  However, for the case in the last column (the entire lower part of the target point set is missing), our method mistakenly aligns the two objects by shifting the source point set higher. For these cases, without additional adjustment to the missing points, minimization of the pre-defined unsupervised alignment loss eventually leads to alignment failures. 

\subsection{Resistance to Data Outliers (D.O.) Noise}\label{sc_do}
In this experiment, we further verify our model's performance for 3D rigid point set registration in the presence of D.O. noise.\\

\begin{figure*}
\begin{center}
\includegraphics[width=17cm]{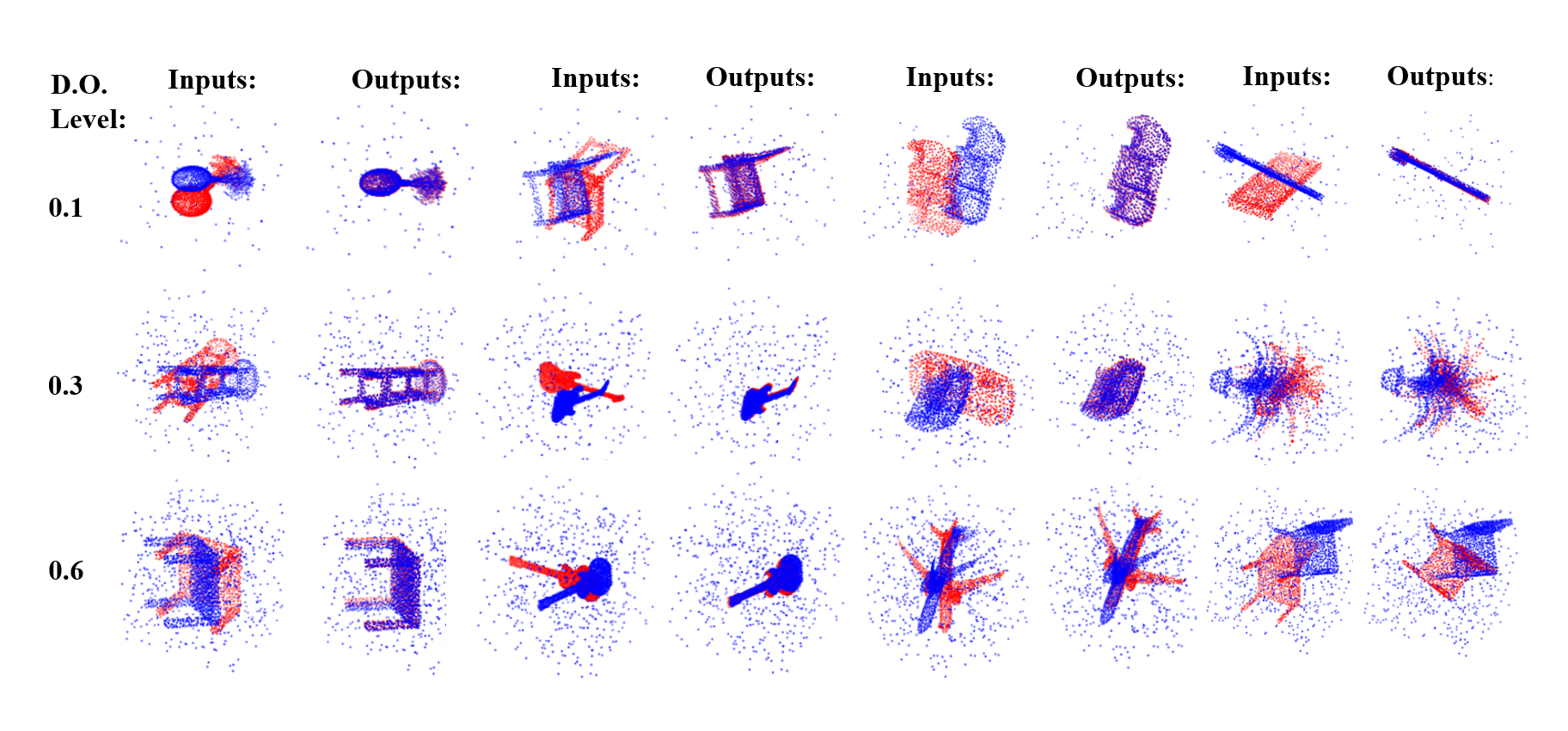}
\end{center}
\caption{Randomly selected qualitative results in presence of D.O. noise. Left columns: inputs. Right columns: outputs. The red points represent source point sets, and the blue points represent the target point sets.}
\label{fig4}
\end{figure*}

\noindent{\textbf{Experiment Setting:}} To test our model's performance for 3D rigid point set registration in the presence of D.O. noise, we firstly split ModelNet40 as explained in section 4.3. Then we add the D.O. noise on the target shape by the way introduced in section 4.1. We compare the performance of our model with DCP model. Both models are trained on the training set the evaluated on the test set. The quantitative result for D.O. noise level 0.1 is demonstrated in Table \ref{tab999} and some randomly selected qualitative results for various D.O. noise levels from 0.1 to 0.6 are shown in Figure \ref{fig4}.  We adjust the Chamfer distance loss by clipping large values which exceed 0.1. \\

\noindent{\textbf{Results:}} As shown in Table \ref{tab999}, for dataset in presence of D.O. noise, the quantitative result achieved by our model is better than the results of DCP. In comparison to the method DCP, for the precision of rotation matrix estimation, our method achieves 16.57 MSE(R) in comparison to 46.99 achieved by DCP. For the precision of translation matrix estimation, our method achieves 0.006 MSE and 0.025 MAE in comparison to 0.002 MSE(R) and 0.037 MAE achieved by DCP. Even though our model is trained in an unsupervised way, we have a comparable performance for translation prediction in the presence of D.O. data noise. As we can be seen in Figure \ref{fig4}, our model performs well when the D.I. level is as low as 0.1, but our model experiences a clear performance drop when D.O. level increased to 0.6. When the D.O. noise increases to 0.6, our model can still align well for approximately half of the cases. Some failures cases may occur due to the initialization problem when our model faces heavy outliers. The cases with D.O. noise needs more focuses and works in future study. Here we only show our model's performance without specially dealing with the outliers.

\begin{table*}
\centering
\small
\begin{tabular}{ccccccc}
\hline
Model & MSE(R)& MAE(R) & MSE(t) & MAE(t)\\
\hline
Ours & 16.5751 & 1.3631 & 0.0060 & 0.0255  \\
DCP & 46.992622 &  4.586546 & 0.002941 & 0.037136  \\

\hline
\end{tabular}
\caption{Quantitative result for 3D point set registration in presence of D.O. noise.}
\label{tab999}
\end{table*}

\begin{table*}
\centering
\small
\begin{tabular}{ccccccccc}
\hline
Methods& MSE(R)& RMSE(R)& MAE(R) & MSE(t) &RMSE(t)& MAE(t)\\
\hline
Ours &1.154405 &1.074432&0.830864&0.000444 &0.020904& 0.014533 \\
Direct Optimiz. & 406.131713&16.454065&13.932246&0.087263&0.295404&0.253658\\
\hline
\end{tabular}
\caption{Quantitative comparison result for 3D point set registration between our model and direct optimization algorithm.}
\label{tab888}
\end{table*}

\subsection{Oblation study: comparison with direct optimization method}\label{sc_direct}
In this experiment, we conduct further experiments to verify the design of our spatial-correlation representation (SCR) and decoder network. \\

\noindent{\textbf{Experiment Setting:}}
To test the effectiveness of our designed decoder with the learned prior knowledge, we compare our model with the direct optimization algorithm. For the direct optimization algorithm version, we optimize the same alignment loss by directly passing the gradients directly to R, t matrix. For preparing the dataset, we follow exactly the settings as section 4.3. The direct optimization algorithm is directly tested on the testing set. Our model is pre-trained on the training dataset without using any label information and then tested on the testing set for comparison. We use the same evaluation metric as in section 4.3 and the quantitative result is demonstrated in Table \ref{tab888}. \\

\noindent{\textbf{Results and Discussion:}} As is shown in Table \ref{tab888}, directly optimizing R,t with regard to the Chamfer loss cannot lead to a feasible solution of the point set registration problem. This method only leads to an MSE(R) of 406.13 for notation prediction, which is unacceptable in real-world scenarios. With the same experimental settings, our proposed method gets significantly better performance, which demonstrates the effectiveness of the proposed SCR representation and decoder network.\\

Given source and target point sets with a size of $N = 2048$, we can first define a relative position tensor $N\times 3$ which characterizes the relative positions between two point sets and the alignment is measured using Chamfer loss. Our approach essentially leverages a multi-layer neural network to learn the nonlinear mapping function that maps the relative position tensor $N\times 3$ to geometric transformation representation (R,t) (dimension of 6). Technically, it is possible to directly optimize the chamfer distance over the parameters (R,t) by setting up one single-layer network (i.e. the geometric transformation representation layer). However, practically it is not feasible to just train a single-layer neural network that is capable of mapping the high-dimensional relative position tensor (e.g. dimension of $3,000$ for point sets of $1,000$) to low-dimensional geometric transformation representation (R,t) (dimension of 6). Therefore, our method uses a multi-layer neural network to model this non-linear mapping/dimension reduction problem. In addition, to better formulate the concept of relative position tensor, we propose to design spatial correlation representation (SCR) features. During training, we jointly optimize the SCR and decoder to complete the non-linear mapping process. In the testing phase, we fix the trained decoder and only optimize the SCR for a new given pair of source and target point sets.

\section{Conclusion}
This paper introduces a novel unsupervised learning-based approach to our research community for point set registration. In contrast to recent learning-based methods (i.e. Deep Closest Point), our Deep-3DAligner gains competitive advantages by eliminating the side effects of the hand-craft design in feature encoder and correlation module. We conducted experiments on the ModelNet40 datasets to validate the performance of our unsupervised Deep-3DAligner for point set registration.  The results demonstrated that our proposed approach achieved comparative performance compared to most recent supervised state-of-the-art approaches. Our Deep-3DAligner also achieves reliable performance for 3D point sets in the presence of Gaussian noise, outliers, and missing points.

\bibliographystyle{IEEEtran}
\bibliography{egbib}

\end{document}